\documentclass[graybox]{svmult}

\usepackage{mathptmx}       
\usepackage{helvet}         
\usepackage{courier}        
\usepackage{type1cm}        
\usepackage{makeidx}         
\usepackage{graphicx}        
\usepackage{multicol}        
\usepackage[bottom]{footmisc}

\newcommand{\be}{\begin{equation}}
\newcommand{\ee}{\end{equation}}
\newcommand{\ba}{\begin{eqnarray}}
\newcommand{\ea}{\end{eqnarray}}
\newcommand{\bc}{\begin{center}}
\newcommand{\ec}{\end{center}}
\newcommand{\lsi}{LS~I~+61$^{\circ}$303}
\newcommand{\ls}{LS 5039}

\makeindex             

\begin{document}

\title*{$\gamma$-ray binaries as non-accreting pulsar systems}

\author{Diego F. Torres}

\institute{Diego F. Torres \at Instituci\'o Catalana de Recerca i Estudis Avan\c{c}ats (ICREA), Catalunya, Spain; and Institut de Ci\`encies de l'Espai (IEEC-CSIC), 
Campus UAB, Fac. de Ci\`encies, Torre C5, parell, 2a planta
08193 Barcelona,  Spain.
\email{dtorres@ieec.uab.es}}

\maketitle

\abstract*{ The $\gamma$-ray  binaries LS 5039 and \lsi\ have been detected by Cerenkov telescopes at TeV energies, exhibiting periodic behavior correlated with the orbital period. These $\gamma$-ray binary systems have also been recently detected by the {\it Fermi} Gamma-ray Telescope at GeV energies, and combination of GeV and TeV observations are providing both, expected and surprising results. 
We summarize these results, also considering the multi-frequency scenario, from the perspective of pulsar systems. We 
discuss similarities and differences of models in which 
pulsar wind/star wind shocks, or pulsar wind zone 
processes lead to particles accelerated enough to emit 
TeV photons. We discuss in detail the caveats of the current observations for detecting either accretion lines or pulsations from these objects. 
We also comment on the possibility for understanding the GeV to TeV emission from these binaries with a 2-components contribution to their spectrum. We show that it would be possible to accommodate both, normal pulsar emission and GeV / TeV fluxes that vary with orbital phase. We point out several aspects of this idea that are subject to test with data being currently taken. }

\abstract{  The $\gamma$-ray  binaries LS 5039 and \lsi\ have been detected by Cerenkov telescopes at TeV energies, exhibiting periodic behavior correlated with the orbital period. These $\gamma$-ray binary systems have also been recently detected by the {\it Fermi} Gamma-ray Telescope at GeV energies, and combination of GeV and TeV observations are providing both, expected and surprising results. 
We summarize these results, also considering the multi-frequency scenario, from the perspective of pulsar systems. We 
discuss similarities and differences of models in which 
pulsar wind/star wind shocks, or pulsar wind zone 
processes lead to particles accelerated enough to emit 
TeV photons. We discuss in detail the caveats of the current observations for detecting either accretion lines or pulsations from these objects. 
We also comment on the possibility for understanding the GeV to TeV emission from these binaries with a 2-components contribution to their spectrum. We show that it would be possible to accommodate both, normal pulsar emission and GeV / TeV fluxes that vary with orbital phase. We point out several aspects of this idea that are subject to test with data being currently taken. }


\section{Prologue}

Four massive binaries have been discovered as variable very-high-energy (VHE) $\gamma$-ray sources. They are PSR B1259-63 (Aharonian et al. 2005a), LS 5039 (Aharonian et al. 2005b, 2006), \lsi\ (Albert et al. 2006, 2009, Acciari 2008,2009), and Cyg X-1 (Albert et al. 2007). HESS J0632+057 was found by the H.E.S.S. experiment (Aharonian et al. 2007) as one of the very few point like sources. Its positional association with the massive star MWC 148 led to the suggestion that its nature is that of a gamma-ray binary system, albeit this has not yet been confirmed (Hinton et al. 2009).
PSR B1259-63 is obviously a system formed with a pulsar whereas Cyg X-1 is most likely formed with a black hole compact object. The nature of the two remaining systems (\lsi, and LS 5039,  recently detected by {\it Fermi}, see below) is not yet settled. Variable gamma-ray emission was also reported from Cyg X-3, but it is only detected at GeV energies (Abdo et al. 2009c, Aleksi\'c et al. 2010).
The phenomenology presented by the systems that have been detected both at TeV and GeV energies can be distinguished in two classes: there are those presenting recurrent TeV emission correlated with the orbit (the case of \lsi, LS 5039, and PSR B1259-63) and there is also one case (Cyg X-1) that was hinted at in a flaring episode that has not been found to repeat yet. Cyg X-1 and 
the  three other TeV sources also differ in their SEDs. 
In Cyg X-1, the transient VHE luminosity was less than 
1\% of the X-ray luminosity. 
This Chapter analyzes some aspects of the high-energy and multi-wavelength phenomenology, especially of \lsi\ and \ls, in what concerns to their possible interpretation as non-accreting pulsar systems.

\begin{figure}[t]
  \begin{center}
   \includegraphics[height=.23\textheight]{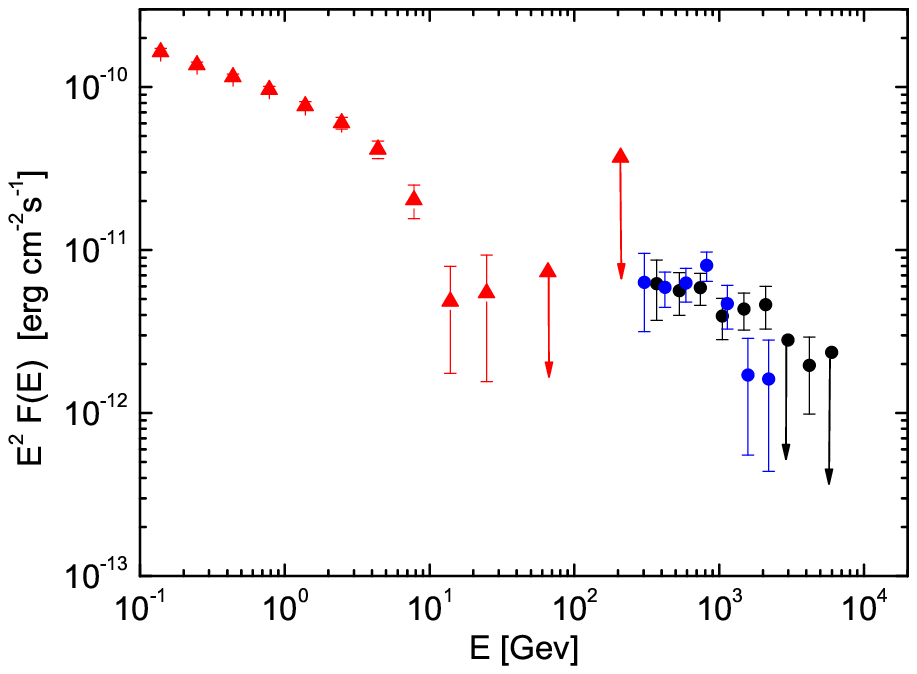}
   \hspace{-0.8cm}
     \includegraphics[height=.23\textheight]{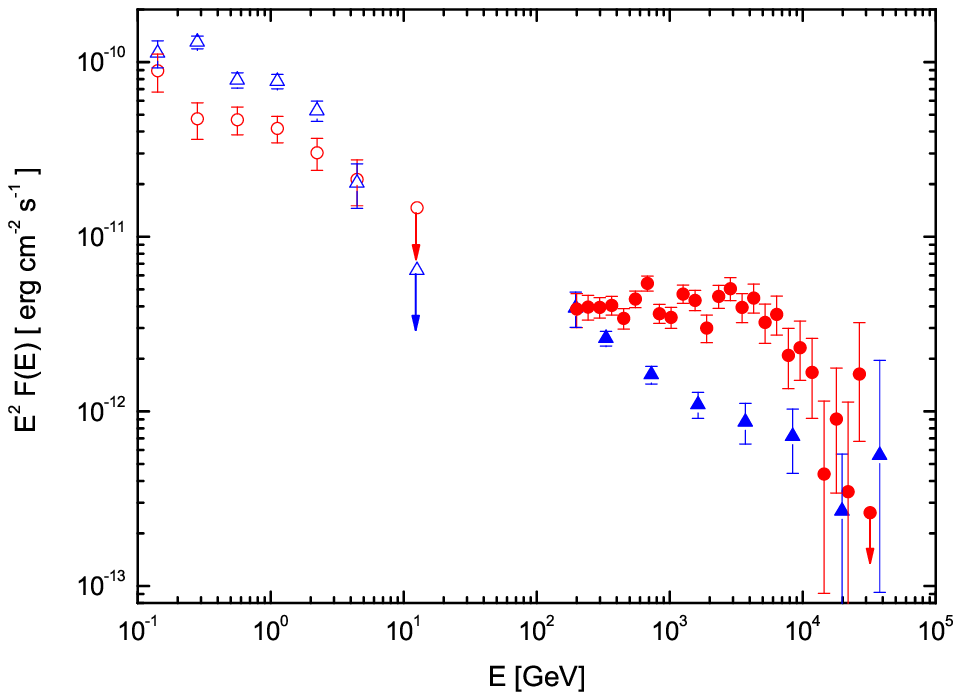}
    \end{center}
  \caption{ Summary of GeV to TeV observations of the two $\gamma$-ray binaries detected with {\it Fermi}.
  High and very-high energy spectral data points of \lsi\ (left) and \ls\ (right). Left: {\it Fermi} (integrating the whole orbit) data points are triangles; MAGIC data are lighter circles  (high state phases 0.5-0.7);  and VERITAS data are black circles (0.5-0.8). 
Right: Circles show the spectral data at 
INFC (TeV maximum, GeV minimum, phases 0.45--0.9; triangles  show the spectral data at SUPC
(TeV minimum, GeV maximum, phases $<0.45$ and $>0.9$). Higher energy data comes from H.E.S.S. 
Note that the data from the different telescopes are not contemporaneous, though they do cover multiple orbital periods.}
  \label{fermiLSspectrum}
\end{figure}

In this same volume, contributions by R. Dubois et al. ({\it Fermi} collaboration) and J. Cortina give account of the observations at high and very-high energies of these sources. We shall only make notice of the spectral energy distribution,  referring  to the work of the former authors for details. 
Figure \ref{fermiLSspectrum} shows the high and VHE spectrum for both LS 5039 and \lsi.
For \lsi, the data plotted at the two energy bands are not corresponding to the same part of the orbit, with the {\it Fermi} data being an average. In the case of LS 5039, two broad band spectra --corresponding to the inferior and superior conjunction phase intervals-- are shown for both energy bands, and although not contemporaneous, they cover multiple orbital periods.

\section{Why is a non-accreting pulsar system a tenable alternative?}

In this section, we put some important aspects of the multi-wavelength information on \ls\ and \lsi\ in context of the highest energy detection. We list the main reasons by which one can in principle sustain that a non-accreting pulsar system is a tenable alternative for these binaries (see also Dubus 2006, and Zdziarski, Neronov \& Chernyakova 2008). \\

 {\it  1) The mass function does not provide definite constraints to the mass of the compact object, with a pulsar being an acceptable (if not preferred) alternative:}\\

Combining Newton's laws of gravitation and motion, the mass function is
$ f(m_1,m_x) = {({4\pi^2} / {G})}{({(a \sin i)^3} / {P_b^2})} = {{(m_x\sin i)^3} / {(m_1+m_x)^2}} $, where
$m_1$ and $m_x$ are the masses of the star and companion respectively, $G$ is Newton's gravitational constant and $i$ is the inclination angle of the binary orbit (defined so $i=90^0$ is edge on). The mass function has then units of mass, and is the minimum mass of the $x$-companion 
corresponding to a zero-mass $m_1=0$ and inclination angle $i=90^0$: $m_x > f(m_1,m_x)$. This means that the determination of the inclination and companion mass is mandatory to determine the type of the compact object. If $m_1 = 10 - 15$ M$_\odot$, and $f(m_1,m_x) \sim 0.01$ the companion would be a neutron star if $i > 25^0$. 
The mass function values found for LS 5039 and \lsi\ (see e.g., Casares et al. 2005a,b; or Aragona et al. 2009) are  low ($\sim$ 0.01). To constrain the compact object mass from the mass function, the orbital inclination and mass of the optical primary must be known, which are subject to their own uncertainties. The parameter space derived by Casares et al. (2005a,b), by Grundstrom et al. (2007), and by Aragona et al (2009)  all allow for both a neutron star or a black hole in LS 5039 and \lsi\, the inclination of the orbit being poorly constrained (limits are not strict) $10^0 < i < 60$, or even higher. It is interesting to note that the assumption of a neutron star in both objects, of 1.4 M$_\odot$ would imply a large inclination: the compact object would be a neutron star if $i>25^0$ and a black hole otherwise. If ignorance of the orbital inclination can be represented by a random distribution of $i$, the likelihood for the system to be formed by a neutron star is greater than if it is formed by a black hole compact object. \\

 {\it 2) There is no clear sign of an on-going accretion process:}\\

The X-ray and radio properties of LS 5039 and \lsi\ distinguish them from other X-ray binaries, and particularly for \lsi, set it apart from other Be X-ray binaries (XRBs) (see below), since they both present no conclusive signs of accretion. If these systems accrete, given the sizes of their orbits, accretion would have to be wind-fed. Periodic ellipsoidal variations are also a common phenomenon among high-mass X-ray binaries (HMXBs), provided that the optical companion fills its Roche lobe,   due to tidal distorsions of the optical star. These are also not observed (Mart\'i et al. 2004).
The Bondi mass accretion rate is $\dot M \sim \dot M_w / (2 r_a / d_s)^2$ where $\dot M_w$ is the stellar wind mass loss rate (typically about 10$^{7-8}$ M$_\odot$ yr$^{-1}$), $d_s$ is the orbital separation, and $r_a = 2 G M_c / v_w^2$ is the Bondi capture radius, with $M_c$ the mass of the compact object and $v_w$ the wind speed. This implies average values, e.g., for LS 5039 and assuming a polar flow of 2000 km s$^{-1}$, of $5 \times 10^{14}$ g s$^{-1}$. This value of mass accretion, when transformed into power, is close to, or even lower than the GeV luminosity of the system, implying unrealistic efficiencies (see, e.g., Dubus 2006).
In order to avoid disc or magnetospheric accretion, the relativistic wind of the pulsar must be able to stop the infall of
stellar matter, implying an upper limit to its putative period. 
Writing the spin-down power
as a function of the pulsar magnetic field $B$ and period $P$, it results 
$P < 230 B^{1/2} \dot M_{15}^{-1/4}$ ms, with $\dot M_{15}$ being the scale $\dot M /10^{15}$ g s$^{-1}$.  Thus, if a fast millisecond pulsar is present in these systems,  accretion onto it is not expected.

As e.g., Zhang et al. (2010) confirmed with the analysis of deep INTEGRAL observations of \lsi, there is no high-energy cut-off at energies below 100 keV, as it would be common in conventional accretion scenarios. Indeed, if the system is an accreting neutron star or black hole, one expects to find a cut-off power- law spectrum in the hard X-ray band with a cut-off energy normally at 10 -- 60 keV for neutron stars (e.g., Filippova et al. 2005) and at $\sim$100 keV for black holes (McClintock \& Remillard 2003). See also the LS 5039 Suzaku observations by Takahashi et al. 2009, where the X-ray spectral data up to 70 keV are described by a hard power-law with a phase-dependent photon index which varies within 1.45--1.61).
Regarding spectral lines due to accretion,   {\it RXTE} spectra of LS 5039 did appear to show a strong, broad Fe line (Rib\'o et al., 1999), although it was not confirmed by other X-ray missions (see, e.g., the  {\it XMM-Newton} observations by Martocchia et al., 2005). Several authors argued that being LS 5039 in the Galactic Plane, the Fe line is very likely due to Ridge emission sampled by the large  {\it RXTE} field of view (e.g., Bosch-Ramon et al., 2005; Dubus 2006a, Zdziarski, Neronov, Chernyakova 2008). We discuss the prospects for line detection from these systems in greater detail below, where we put forward a note of caution given the relatively limited observation times that have been granted compared with the amount of time needed to detect accretion lines from these systems under some assumptions. \\

{\it 3) If there is a pulsar, non-detection of radio pulses is expected:}\\

Zdziarski, Neronov \& Chernyakova (2008) have already shown in detail why radio pulses would be absorbed in these compact binaries. We follow their derivation in what follows:
The free-free absorption coefficient due to ions with the atomic charge, $Z$, is given by 
\be
\alpha_{\rm ff}= (2^{5/2} \pi^{1/2} e^6)/( 3^{3/2} m_{\rm e}^{3/2} c)
(kT)^{-3/2} Z^2 n_{\rm e} n_{\rm Z} \nu^{-2} \bar g,
\ee
where $\nu$ is the frequency, $n_{\rm Z}$ is the $Z$-ion density, $n_{\rm e}$ is the electron density, and $\bar g$ is the average Gaunt factor.\footnote{The Gaunt factor for $h\nu\ll kT$ and $\nu \gg \nu_{\rm p}$ (where $\nu_{\rm p}$ is the plasma frequency) equals
$
\bar g =(3^{1/2}/ \pi)  [ \ln ( (2kT)^{3/2} ) / ( \pi e^2 Z
 m_{\rm e}\nu)) - 5\gamma_{\rm E} /  2 ],
$
where  $\gamma_{\rm E}\simeq 0.5772$ is Euler's constant.} Using $Z=1$, $T= 10^5$ K and $\nu=5$ GHz, and
averaging over $Z$, 
$
\alpha_{\rm ff} \simeq 0.12 T^{-3/2} ({\mu_{\rm i}^2 }/{ \mu_{\rm e}} )
n_{\rm i}^2 \nu^{-2}\, {\rm cm}^{-1}\simeq 0.175 T^{-3/2} n_{\rm i}^2 \nu^{-2}\, 
{\rm cm}^{-1},
$
where $T$ and $\nu$ are in units of K and Hz, respectively, and $\mu_{\rm e}=2/(1+X)\simeq 1.2$ is the mean electron molecular weight.
For the equatorial disc, the density can be expressed as 
$
n_{\rm d,i}(D)\simeq n_{\rm d,0} (D  / R_\star )^{-\gamma},
$
where $n_{\rm d,0}\sim 10^{13}$ cm$^{-3}$, $\gamma\simeq 3.2$, and $D$ is the  distance from the center of the Be star (see, e.g., Waters et al. 1988, and the discussion below).
The optical depth perpendicular to the disc plane is subject to a disc thickness of $2D\tan \theta_0$, with $\theta_0$ the half-inclination angle. Thus, 
\begin{equation}
 \frac{\tau_{\rm d,ff} }{6 \times 10^6 } \simeq
  \left[n_{w,0}\over  
10^{13}{\rm cm}^{-3}\right]^2 \left[\frac{T}{10^5\mbox{K}}\right]^{-3/ 2} \left[\frac{\nu}{1 \mbox{GHz}}\right]^{-2} \left[\frac{D}{3\times 
10^{12} \mbox{cm}}\right]^{-5.4}
\end{equation}
where $3\times 10^{12}$ cm corresponds to the periastron separation of \lsi.
Thus,  the equatorial disc is optically thick to radio emission, including of course, pulsations. This is consistent with the suppression of pulsed radio emission observed close to periastron in the system PSR B1259-63 (Johnston et al. 1992, Melatos et al. 1995).
To calculate the radial optical depth of the fast polar wind from infinity down to a given value of $D$
we can consider a clumpy wind (free-free absorption would still provide a high opacity even for smooth polar flows) where the wind density inside the clumps is $1/ f$ times that of the smooth wind, 
$\langle n_{\rm i}\rangle \simeq n_{\rm i} f$ and
\begin{equation}
\frac{ n_{\rm w,i}}  {3\times 10^{8} {\rm cm}^{-3}} \simeq \left[\frac{v_\infty}{10^8\mbox{cm s}^{-1}} \, {f\over 0.1}\right]^{-1}  
\left[\frac{\dot M_{\rm w}}{10^{-8}M_\odot \mbox{yr}^{-1}}\right]
\left[\frac{D}{3\times 10^{12}\mbox{cm}}\right]^{-2} .
\end{equation} 
In a clumpy medium, $\tau_{\rm ff}$ is an integral over the square of the density within the clumps times $f$, i.e., $n_{\rm i}^2 f$ (or equivalently, an integral over $\langle n_{\rm i}\rangle^2 f^{-1}$). 
The opacity then results,
\begin{eqnarray}
\frac{ \tau_{\rm w,ff} } {5\times 10^3}   &\simeq&  \left[\frac{\dot M_{\rm w}}{10^{-8} M_\odot  \mbox{yr}^{-1}}\right]^2 \left[\frac{v_\infty}{10^8\mbox{cm s}^{-1}}\right]^{-2} \left[f\over 0.1\right]^{-1} \nonumber\\ &\times&
\left[ \frac{\nu}{1 \mbox{GHz}} 
\right]^{-2} \left[\frac{T}{10^5\mbox{K}}\right]^{-3/2}
\left[\frac{D}{3\times 10^{12} \mbox{cm}}\right]^{-3}.
\end{eqnarray}
%
Thus, according to the orbital solutions found for the systems, the neutron star would move in the optically thick region, explaining the absence of observed radio pulsations. In addition of the high opacity, one can always entertain the possibility for the radio emission cone to be emitted off our line of sight, thus missing it entirely as in the case of any undetected-in-radio, bright GeV pulsars, even if at some portion of the orbit, radio pulses could escape. Pulsed emission in the X-ray or GeV band would  be as important as those in radio to pinpoint a pulsar component, and we come back to these possibilities below.\\

{\it 4) There are no clear signs of jets and the radio morphology is consistent with a pulsar component:}\\

From $\sim$50 mas resolution radio images of \lsi\ obtained
with Multi-Element Radio Linked
Interferometer (MERLIN), extended, apparently precessing, radio emitting structures at angular extensions of $0.01-0.05$ arcsec have been reported by Massi et al. (2001, 2004). 
However, recent Very Long Baseline Array (VLBA)  imaging obtained by
Dhawan et al. (2006) over a full orbit of \lsi\ has shown the radio emission to
come from angular scales smaller than about 7~mas (which is a projected size of 14 
AU at an assumed distance of 2~kpc).
This radio emission
appeared cometary-like, and interpreted to be pointing away from the high
mass star and thus being the smoking gun of a pulsar wind. A careful analysis of the images and phases, however shows that this interpretation is not straightforward: The {\em tail} is not always seemingly pointing in the {\em right} direction, at least when the polar flow of the Be star is considered.
Nevertheless, there are some  
unambiguous results of these observations: a) No large features or high-velocity flows were noted in any of 
the observing days, which implies at least its non-permanent nature. b) The changes within 3 hours were found to be insignificant, so the velocity 
of the outflow can not be much over 0.05$c$.

\begin{figure} [t]
    \includegraphics[height=2.9 cm]{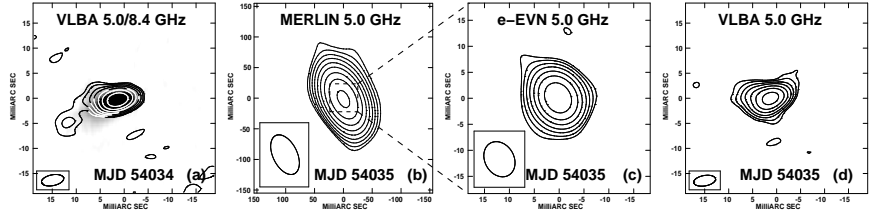}
\caption{Radio images of LS~I~+61\,303\, 
obtained on 25 October 2006 with the VLBA (panel a) and on 26 October
2006 with MERLIN (panel b), EVN (panel c), and VLBA (panel d). 
In panel a, the 8 GHz VLBA image on 2 February 2006 (grey scale) is also overlaid, convolved with the 5 GHz
VLBA beam. This date corresponds to the same phase of LS~I~+61\,303\,
($\phi_\textrm{orb} \approx$0.62). In all cases, the origin of coordinates is set at the VLBA
peak of brightness on 25 October 2006, and the contours are drawn at
(3,3\,$\sqrt{3}$,9,...) times the off-source rms. Figure reproduced from Albert et al. (2008), with permission of AAS. }
\label{fig:radio:magic}
\end{figure}

The MAGIC collaboration also conducted a radio campaign (in concurrence with TeV observations) to test these results using again   MERLIN in the UK, the European VLBI Network (EVN),
and the VLBA in the USA (Albert et al. 2008). In this campaign, radio observations at different angular resolutions were conducted at the same time. 
It is then interesting to focus on the day where all these facilities observed at the same time (October 25-26, 2006). 
The results obtained by radio imaging at different angular scales show
that the size of the radio emitting region of \lsi\ is constrained to be
below $\sim$6~mas ($\sim$12 projected AU), and the presence of
persistent jets above this scale is therefore excluded. 
As in the case of Dhawan et al. (2006), these observations have shown a radio-emitting region 
extending east-southeast from the brighter, unresolved emitting core. The
outflow velocity implied by these observations is 
$\sim 0.1c$. The comparison between Dhawan  et al. (2006) and
Albert et al. (2008) images at the same orbital phase (but obtained 10 orbital cycles apart) show a high
degree of similarity on both its 
morphology and
flux, which suggests periodicity and stability of the physical processes involved in the
radio emission. These would be hard to attain if, for instance, the radio emission is the result of the random interaction of a steady flow with wind clumps. On the other hand, 
if the
radio emission is produced by a milli-arcsecond scale jet, the required
stability and periodic behavior of such a jet in order to produce the same radio map across many orbital cycles 
would be difficult to
reconcile with the non-persistent nature of a larger scale ($\sim$100
mas) relativistic jet.  Figure \ref{fig:radio:magic} shows that  no jet can be seen at any scale.

A similarly changing milli-arcsecond radio morphology was found for LS 5039 (Rib\'o et al. 2008). 
In fact, early detection of elongated asymmetric emission in high-resolution radio images obtained with VLBA and EVN was interpreted as evidence of its microquasar nature, and suggested that the source was persistently producing jets with mildly relativistic ejections with a velocity of $\sim 0.15c$ (Paredes et al. 2000, 2002). Recent analysis by Rib\'o et al. (2008) show that two images obtained five days apart (recall the orbital period of about 4 days in this system, implying that the observations occurred in phases $\phi = 0.43-0.51$  and $0.71-0.79$) present a changing morphology. There is a core component with a constant flux density, and an elongated emission with a position angle (PA) that changes by $12^0 \pm 3^0$, with the source being nearly symmetric in the first run and asymmetric in the second one. Analysis of the changes in the position angles, the inferred velocities of the outflows, and the steadiness of the radio flux, all make for an unpreferred interpretation in the framework of a microquasar model. As the authors conclude, they more naturally point to a young non-accreting pulsar scenario where the different morphologies detected at different orbital phases are due to the change of the relative positions between a pulsar and the companion star along the orbit.

Dubus (2006a) has earlier explored this morphology change in pulsar wind/stellar wind scenarios, predicting a periodic change in the direction and shape of the extended radio morphology as well as in the peak position of the radio core, depending on the orbital phase. 
The key concept is that whereas on 
a small scale, the shocked material flows away from the binary on a straight path, following the direction given by the vector difference of the stellar wind and orbital speeds $\mathbf{v}_{\rm w}-\mathbf{v}_{\rm orb}$; on a larger scale, the material shears with the orbital motion becoming important at $\sim d_{\rm n} \sim \sigma c P_{\rm orb} / 2\pi$ (about 1 AU for LS 5039).\footnote{This scale results from a comparison between the relevant flow timescale $\tau_{\rm flow}=d/v$ to reach a distance $d$ from the shock and the orbital motion timescale $\tau_{\rm orb}=d_s/v_{\rm orb}$ which can vary a lot for a highly eccentric orbit, and having taken $v$ equal to its asymptotic value $\sigma c$ ($\sigma$ is the ratio of magnetic to kinetic energy, $\sigma = (B^2/4 \pi) / (\Gamma nm_e c^2))$. Thus, the shape and location of the cometary outflow can ultimately be used to constrain the magnetization. } These effects combine to provide non-trivially shaped radio-maps, similar in aspect to those observed. The maps at $\phi$=0-0.25 are similar to the radio-morphology observed by Paredes et al. (2000), see Dubus (2006a). \\

{\it 5) A young age of the system is consistent with observations:} \\

If the compact object is a young and energetic millisecond pulsar (necessary to avoid accretion onto it), the system should then be similarly young. Rib\'o et al. (2002) have put  an upper limit to the age of LS 5039 by  tracing back the proper motion of the system to the plane.  LS 5039 is a runaway system, with a total systemic velocity of about 150 km s$^{ -1}$ and a component perpendicular to the galactic plane larger than 100 km s$^{ -1}$. This is probably the result of an acceleration obtained during the supernova event that created the compact object in this binary system. Nitrogen enrichment in the atmosphere of the companion star also suggests a young age in LS 5039 (McSwain et al. 2004; rotational-mixing could also explain the
enrichment, Casares et al. 2005b, see Dubus 2006). Note too that radio searches of supernova remnants have also failed for PSR B1259-63, which has a measured spin-down age of $3 \times 10^5$ years (Johnston et al. 1992). \\

{\it 6) Pulsar systems would be consistent with population predictions of Be XRBs:}\\

The Be XRBs, like \lsi\ are the most numerous class of XRBs known. 
At present, 64 Be XRBs are known in the Galaxy, and in 42 the compact object was confirmed to be a neutron star (NS) by the presence of the X-ray pulsations; in not a single one of them, a black hole was confirmed (Belczynski \& Ziolkowski 2009). Recent population synthesis made by the latter authors are consistent with 0-2 black hole Be XRB in the Galaxy, which can perhaps be understood as a result of an evolutionary effect. The  rotation of Be stars could be achieved during a period of Roche-lobe overflow mass transfer from its initially more massive companion, and if it looses most of the mass in the early pre-SN phase process, becoming a He star with mass of only a few M$_\odot$, when exploding as a supernova it can only leave a neutron star behind (see e.g., Tauris \& van den Heuvel 2006).
One has to bear in mind, however, that most of these Be XRBs are accreting systems, hosting pulsars with long periods (older spin-down systems) and only a few are expected to have a young age at any given moment. Spectral-wise, then, e.g., \lsi\ is very different from most of these objects above 10 keV, with the latter being cut off in hard X-rays 
(typically with $\sim$20 keV bremsstrahlung-like spectra) with only very weak emission above $\sim$100 keV as measured, e.g., by  {\it INTEGRAL}.

\section{Caveats in the search for X-ray spectral lines }

\begin{figure}[t]
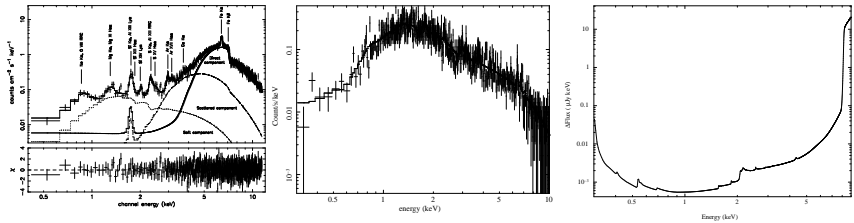

  \begin{center}
    \includegraphics[height=.18\textheight,angle=-90]{4u1700_phase0.15.ps}
    \includegraphics[height=.2\textheight,angle=-90]{model_eclipse_simul_ls5039_10ks.ps}
    \includegraphics[height=.2\textheight,angle=-90]{insensitivity_plot.ps}
    \end{center}
  \caption{Left panel: {\it  XMM} spectrum of the 4-day binary system 4U\,1700-37 at 0.15 of phase (this is a reanalysis of observations made by van der Meer et al. 2005).  Middle panel: simulation of what {\it XMM-Newton} would have seen of the 4U\,1700-37's lines with the available short observations of LS 5039. Right panel: sensitivity plot for the presence of spectral lines in a $\sim$100\,ks long {\it Chandra} observation (the right two panels are adapted from Rea et al. 2010).}
  \label{lines}
\end{figure}

The apparent absence of  spectral lines in the X-ray spectra of TeV binaries has been often used as a proof of their non-accreting nature. However, a word of caution is useful in this respect. 
The X-ray continuum spectra of accretion dominated HMXBs are often described by a power law
with photon index $\alpha \sim 1-2$ (modified at higher energies by an
exponential cutoff between $\sim$30-100\,keV). A spectrum of this form can be produced by
inverse Compton scattering of soft X-rays by hot electrons in the
accretion column near the compact object, and a part of this emission
is scattered by the stellar wind of the massive companion. This
results in a further non-thermal spectral component, but with a
different absorption column depending on the orbital phase of the
system. Furthermore, in some HMXBs a soft excess at $\sim 0.1-2$ keV
is detected, very common in systems hosting pulsars (Hickox
et~al~2004). In luminous systems this soft component can be explained by
reprocessing of hard X-rays from the neutron star by optically thick,
accreting material. For less luminous sources the soft excess is
probably due to other processes, e.g. emission from photoionised or
collisionally heated diffuse gas or thermal emission from the surface
of the neutron star. On top of the continuum model, several spectral
lines are usually present in these systems, neutral and ionized, such
as Fe, Si, Mg, Al, N, Ca, mainly produced in the stellar wind or in the
accretion disk (if any) illuminated by the strong X-ray emission of
the compact object. One clear example of such a system, showing all
the above mentioned spectral components is the HMXB 4U1700-37 (composed by
an O type star in a $\sim$4\,days orbit around a compact object of
unknown nature: see e.g. Clark et al.~2002; van der Meer et
al.~2005; see e.g. Figure\,\ref{lines}).  Note that in 4U1700-37  as well as many other binaries, the presence of spectral lines is highly dependent on the continuum spectrum and on the orbital phase of the system.

So far a detailed high-resolution spectral analysis  has been missing for the two debated TeV binaries: LS 5039 and \lsi. The best data available for these kind of studies come from the {\it XMM-Newton} satellite, thanks to its large collecting area, spectral resolution, as well as the availability of its grating spectrometer. However, at a given orbital phase, only very short observations have been taken (mainly aiming at monitoring the continuum spectral variability over the orbits). In the past {\it XMM-Newton} spectra of, e.g, LS 5039, an EW $\sim 60$ eV is the current 1$\sigma$ limit on the presence of Fe K$\alpha$ (and only in a small part of the orbit), while  in the available {\it Suzaku} observations, the limit on the detection of lines is 40 eV (see Takahashi et al. 2009). Furthermore, with the current short pointings (at a given orbital phase)  there are not enough counts to use anyhow the high-resolution spectral capabilities of the grating cameras, making impossible to answer any question about the presence of narrow lines.

To understand what does this mean, one can consider the binary 4U\,1700-37, a very similar HMXB to LS 5039 (i.e., similar companion star and orbital period, but no TeV emission, and clearly accreting) located at $\sim 1.5$ kpc rather than 3 kpc. Fig. \ref{lines} shows the {\it XMM-Newton} spectrum of 4U\,1700-37 during an eclipse (van der Meer et al. 2005), and how would these accretion lines look, assuming all are present, in the available data for LS 5039. As a result of this simulation, one can see that none of the lines present in 4U\,1700-37 would be detectable by the current data. In Figure \ref{lines} we also show the limits for the presence of lines derived from a 95\,ks long {\it Chandra} recent observation of \lsi\ (Rea et al. 2010).


\section{Caveats in the search for pulsations}

The detection of pulsations is the only unambiguous tracer for a
secure determination of the pulsar nature of the compact
objects hosted in LS 5039 and \lsi. Deep searches for pulsations have been performed in the radio band at several frequencies, with the hope of detecting a fast spinning radio pulsar as in the case of the other TeV binary PSR\,B1259--63 (Johnston et al. 1999, 2005). However, no radio pulsation have detected so far from any of these two sources. As it has been discussed earlier, this is anyway not surprising. In particular, note that at periastron, PSR\,B1259--63 does not show radio pulsations, and that its periastron (given the large orbit, 3.4 year period) has about the same dimension of the major axis of the orbit of \lsi\   and it is way larger than that of LS\,5039's.
On the other hand, searches for pulsations in the X-ray band have many more chances of success than in
the radio band. In fact, the X-ray pulsar beam is usually larger than the radio one, and the strong companion wind does not influence much the X--ray pulsed emission if present. However, what limits the X-ray pulsation search is the pulsed fraction sensitivity that current instruments can achieve. 

Until not long ago, archival observations which could give reliable upper limits on pulsations for fast spinning pulsars ($P \leq 100$\,ms) hosted in LS\,5039 and \lsi\ were not very constraining, coming mainly from {\it R}XTE\, and {\it XMM-Newton} observations. In particular the high background of these instruments (especially {\it R}XTE) limited the pulsed fraction sensitivity of these observations. For \lsi\ the deepest pulsed fraction limit was derived from a 41\,ks {\it XMM-Newton} observation (Sidoli et al.~2006), which was $<28$\% (if not otherwise specified, all pulsed fraction limits here are
calculated assuming a sinusoidal profile and are reported at 90\% confidence level in the 2--10\,keV energy range), in the 12--200\,ms period range.\footnote{Note that the {\it R}XTE\, monitoring observations performed in 1996 gave an upper limit of only 32\% in the 1--200\,ms range, in fact, even though {\it R}XTE's timing resolution and collecting area was much larger than {\it XMM-Newton}, the much higher background contamination is a killer for detecting weak signals, resulting in a larger pulsed fraction limit. For that observations, Harrison et al. (2000) claims a limiting pulsed fraction of $\sim$6\%. However they considered the total count rate without correcting for the cosmic and instrumental background, which if corrected increases substantially the upper limit on the detectable pulsed fraction.} Similarly, for LS\,5039 the deepest limits for the presence of a fast pulsar were derived from a 50\,ks {\it R}XTE\, observation performed in 2003, giving  an upper limit of $P_f<30$\%.  Very recently, to increase the pulsed fraction sensitivity reducing the background contamination, long {\it Chandra} observations have been performed for both  \lsi\ and LS\,5039 (Rea et al. 2010). The results of these observations in the case of \lsi\ are summarized in very deep limits on the presence of X-ray pulsations, with an average limit of $< 10$\% in the 6\,ms -- 10\,s period range (Rea et al. 2010). 
Isolated rotational-powered pulsars emitting pulsed X-rays  are all characterized by  X-ray pulsed fraction much larger than the limits derived by {\it Chandra} for \lsi. However, the presence of the strong stellar wind of the companion, and in particular the shock between the possible pulsar and the stellar wind, is a source of strong X-ray emission.  This unpulsed X-ray emission coming from the shock can be responsible in diluting the pulsed X-ray emission proper to the pulsar, which then might easily end up to represent only $<10$\% of the total X-rays emitted by the systems. Of course, geometry and beaming can also be claimed for missing the pulsations at all.

The possible detection of pulses could also come from observations of the {\it Fermi} telescope.
Blind search pulsations of faint $\gamma$-ray sources require very long observation times, which implies the
calculation of very large fast Fourier transforms (FFT). 
In addition,
the significant frequency derivatives ($\dot{f}$) typical of the $\gamma$-ray pulsars
require that the FFT be repeated many times over a scan of $\dot{f}$.
Specifically, in order to keep the signal power within a single bin of
the FFT, the frequency step should be $\Delta f = 1/T_{obs}$, where
$T_{obs}$ is the total observation time, and the steps size of
$\dot{f}$ would have to be no larger than $1/T_{obs}^{2}$, which means
an enormous number of $\dot{f}$ trials for observation periods as long
as $\sim$1 year.
Atwood et al. (2006) proposed the time-differencing
technique, that reduces the number of $\dot{f}$ trials with a modest
reduction in sensitivity. This method is based on the application of the
FFT on the differences of the photon arrival times, rather than on the
time series itself. In order to be efficient, only the time
differences shorter than a predefined time window $T_w$, that is
significantly shorter than the whole observation, are considered in
the FFT. In this way the number of required steps in $\dot{f}$ is
reduced by a factor $T_w/T_{obs}$.
But unless 
the binary parameters are exactly known, they should also be scanned (like the
$\dot{f}$) to search for a solution. 

The correction of the photon arrival time series from the
solar system reference frame to the pulsar reference frame is mostly
affected by the Roemer delay in the binary system. To evaluate it,
five binary parameters are necessary: the orbital period
($P_{B}$), the epoch of the periastron ($T0$), the argument of the
periapsis ($\omega$), the projected semi-major axis ($A1$),  and the
eccentricity ($e$).
For \ls\ and \lsi, the first two parameters are know with relatively good
precision, but the uncertainties on $A1$, $e$, and $\omega$ are larger than a few percent (e.g, see Casares et al. 2005; worse for \lsi, see Aragona et al. 2009 and references therein).
Thus, without better knowledge of the binary system parameters,
the computational difficulties in running a blind search make the detection of $\gamma$-ray pulsations very challenging.

\section{Notes on the theoretical models based on pulsar systems}

High energy emission from pulsar binaries have been subject of study for a long time (e.g., Bignami et al. 1977, Maraschi and Treves 1981; Protheroe and Stanev 1987, Arons and Tavani 1993, 1994;  Tavani \& Arons 1997; Bednarek 1997; Kirk et al. 1999, Ball and Kirk 2000; Dubus 2006a,b; Sierpowska-Bartosik 2008a,b and others). In most of these studies high-energy primaries are assumed to be accelerated at the shock formed by the collision of the (sometimes, putative) pulsar and the massive star winds. This generic feature allows them to be classified as wind-wind or inter-winds models. In other works the
initial injection is assumed to come directly from the pulsar (the interacting particle population can be a result of equilibrium between this injected distribution and the losses to which it is subject, just as in the case of shock-provided electron primaries; and can also be the result of top-of-magnetosphere or inner-pulsar-wind shocks). In this case, and when opacities to  $\gamma$-ray  production are high due to geometrical and physical characteristics, 
the pulsar wind zone (PWZ) could be the main origin of the high energy radiation, and we can classify this set as intra-wind or PWZ models. 
The closer the binary, the more these two scenarios are expected to produce similar results.  
Almost an order of magnitude difference in opacities can be seen between both systems discussed, with those in LS 5039 being higher, corresponding to the much shorter orbital periodicity (e.g., see Sierpowska-Bartosik \& Torres 2009 for a comparison plot).  Put otherwise, it would imply that a) intra-wind processes can not be completely neglected  for LS 5039, since the opacities for electrons in there is high and cascades can happen  b) inter-wind models can not be completely neglected for \lsi\ (for these kind of more elongated systems, the probability for cascading in the pulsar wind region are lower and most of the energy carried by $e^\pm$ in the PWZ can then be released in the shock region, where the local magnetic field traps the pairs to produce photons via synchrotron and IC processes). These models do not consider either possible contributions of a proton component (which can generate $\gamma$-rays through pp or p$\gamma$ processes (see Chenyakova et al. 2006). These extra components can, as discussed by these authors, also generate additional variability in the electron injection. These models do not contain a relativistic MHD approach of the collision of winds, but are rather based on simplifying assumptions of the hydrodynamical balance (see Romero et al. 2007 and Zdziarski et al. 2008 for a discussion on this issue).

Through the study of intra-winds models, $\gamma$-ray astronomy opens a window to study the electron distribution and magnetization properties of the pulsar winds in binaries, a virgin territory otherwise. Assuming intra-wind models, one can already prove that a mono-energetic distribution of leptons in the putative pulsar wind is ruled out by observations of \lsi\ and LS 5039. Assuming power-law distributions in the winds of pulsars seems justified and possible, although there is a not an priori expectation of the normalization and slope for them, and they so become free parameters of the models, which may in addition be subject to orbital variability, perhaps especially in close binaries. Although more complete models can certainly reduce/erase the need for such a change in the slope of the injection of leptons, as discussed, this is a caveat in the current versions.

In any of these flavors of models, orbital periodicity is naturally encompassed. 
Short and random timescale variability can additionally be expected on top of this periodic behavior, for instance, as a result of granularity or clumping in the stellar wind, or the appearance of random shocks in the inner wind of the pulsar, which may modify local conditions off the average analysis discussed. 
The anti-correlation between GeV and TeV results is naturally encompassed in both of these flavors too (this is resulting just from the opacities evolution along the orbit and the system's geometry, this is a general output of binary models where the radiation is produced close to star even in the case of black holes). IC emission is enhanced (reduced) when the highly relativistic electrons seen by the observer encounter the seed photons head-on (rear-on) i.e. at superior (inferior) conjunction. Inversely, VHE absorption due to pair production will be maximum (minimum) at superior (inferior) conjunction. Gamma rays emitted in the vicinity of the compact object with energies above the $\sim $30 GeV threshold inevitably pair produce with stellar photons (see e.g. Protheroe \& Stanev 1987). On the contrary, emission in the {\it Fermi} range is largely unaffected by absorption but can be affected by cascading of higher-energy photons. The phases of minimum and maximum flux in {\it Fermi} and H.E.S.S., as well as the anti-correlation, are then consistent with these generic expectations, suggesting Inverse Compton scattering is the dominant radiative process above 100 MeV with the additional effect of pair production (affecting photons above 30 GeV) further modulating the  $\gamma$-ray  phenomenology (Bednarek 2006, 2007; Sierpowska-Bartosik \& Torres 2008ab; Dubus et al. 2008).

All in all, these models seem to work reasonably well for the VHE results,
but  both flavors fail in predicting the cutoff found by {\it Fermi} at GeV energies. Browsing SEDs from nominal models published smoothly connect both energy regions and predict levels of GeV-fluxes in excess of what is observed by {\it Fermi} (in the range $\sim$10-100 GeV). The {\it Fermi} cutoff  is an unexpected feature  in  published models and it constitutes an observational surprise.

\subsection{A perspective on the GeV cutoffs of \lsi\ and \ls}

The passage of a compact object through a dense equatorial disk, such as the one commonly found in Be stars, would crush a putative pulsar wind nebula closer to the neutron star, increasing synchrotron losses and introducing a strong dependence with orbital phase of the electron energy distribution (e.g., Dubus 2006b). This may explain the appearance of the cutoff in the {\it Fermi} spectrum of \lsi.
However, there is no such disk around the O6.5V star in \ls. Thus, the existence of the cutoff in both systems argues against explanations related with the properties of the companion, and it seems to require that the radiative process, or the primary injected population, are different in the low and high energy domains (Abdo et al. 2009b). 

The GeV variability is also intriguing. If the GeV 
energy cutoff is interpreted in the outer gap model of pulsars, it is determined by the balance between acceleration and losses to curvature radiation, with the $\gamma$-ray emission being pulsed. The absence of pulsations in {\it Fermi} data is not constraining, given the difficulty of detecting faint pulsars directly in gamma-rays, especially when in binary systems. But even when detected as DC radiation, GeV magnetospheric emission  due to curvature radiation would be produced near the pulsar, and has no obvious reason to be modulated with the orbital motion. In addition, the GeV--TeV anti-correlation would also remain unexplained. 
Despite of this, the orbital-average spectral results for \lsi\, and those found at the SUPC of \ls\ do resemble the spectral properties of pulsars. The typical {\it Fermi} pulsar emission (of which dozens have been measured) has a hard power-law spectrum with a photon index in the range 1.0--2.0 and all of them present an exponential cutoff, the typical value of which is at about 2 GeV (see e.g. the {\it Fermi} pulsar catalog; Abdo et al. 2010).  

\begin{figure*}[t]
\centering
    \includegraphics[height=.25\textheight]{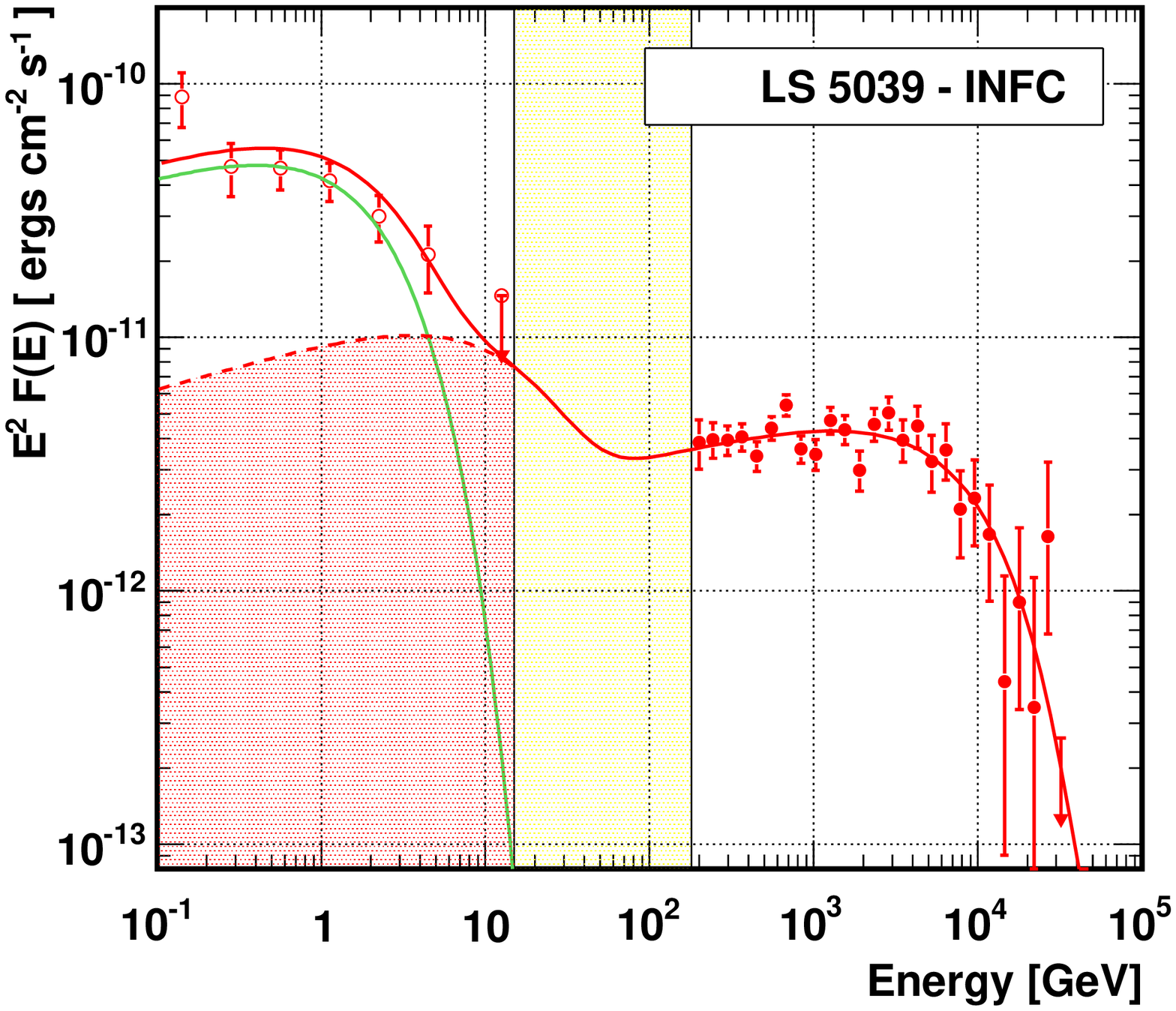}
    \includegraphics[height=.25\textheight]{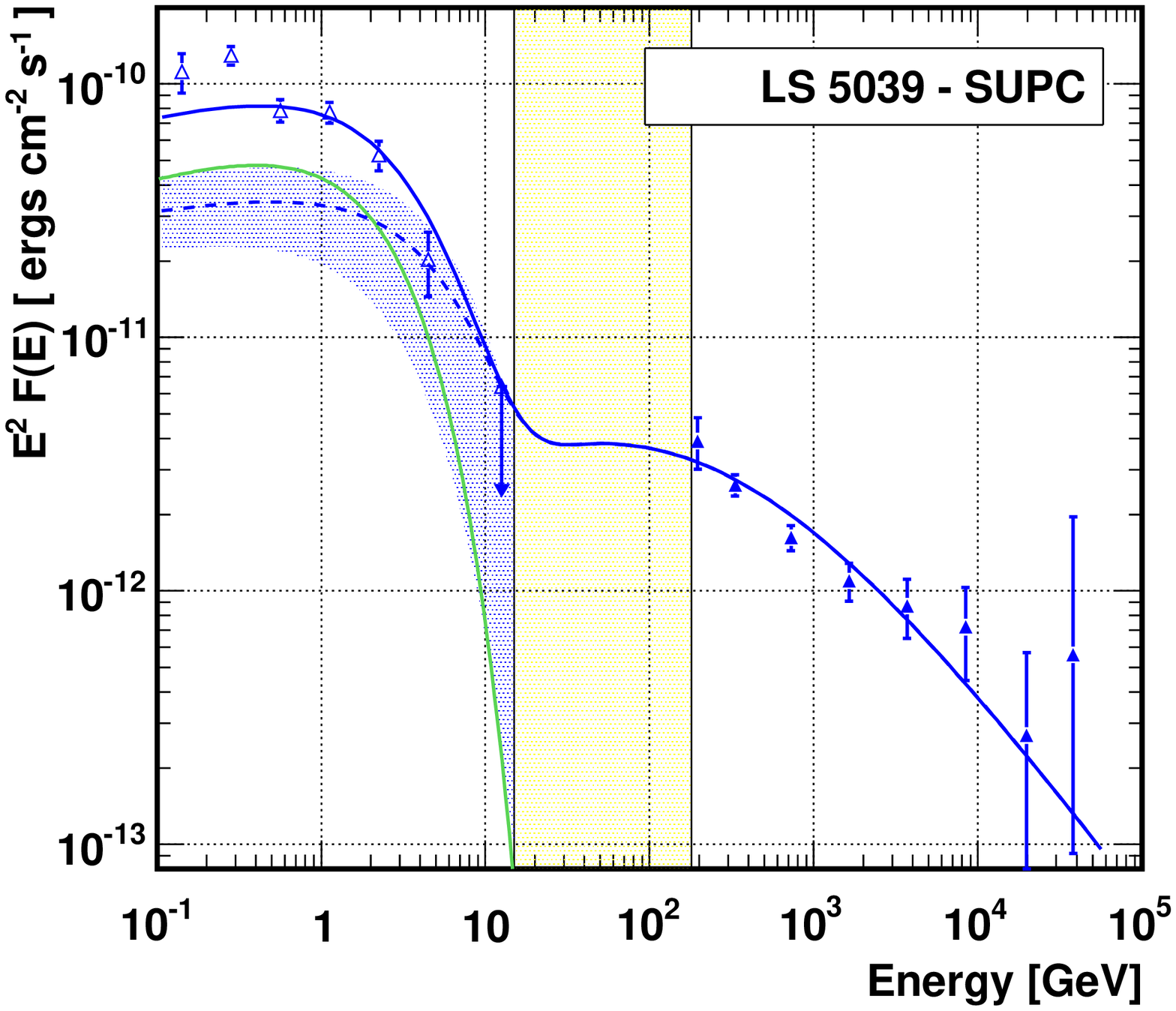} 
  \caption{Interpreting the {\it Fermi} and TeV results of  $\gamma$-ray  binaries with the 2-components idea. See text for details.}
  \label{pulsarinterp}
 \end{figure*}

In what follows, we entertain the possibility that 
the GeV emission detected by {\it Fermi} is produced by two components. One would be the magnetospheric GeV emission coming from a (putative) pulsar in the system. As discussed, this emission is expected to be steady along the orbit (i.e., unaffected by the orbital motion, since its origin is within the pulsar's light cylinder), and pulsed. The other component would come from regions located farther away from the pulsar, i.e., from either the inter-wind region or from the wind zone or from both:  this radiation would then be unpulsed, and it is naturally expected to vary with the orbital phase.  For clarity, we will refer to this un-pulsed component as the wind contribution.

Fig. \ref{pulsarinterp} graphically shows this idea for the richest GeV and TeV sets of data, corresponding to the observations of \ls\ at INFC (TeV maximum, GeV minimum, phases 0.45--0.9; Fig. \ref{pulsarinterp} left panel) and SUPC (TeV minimum, GeV maximum, phases $<0.45$ and $>0.9$; Fig. \ref{pulsarinterp} right panel). Data comes from {\it Fermi} and H.E.S.S. in the GeV and TeV energy ranges, respectively (Abdo et al. 2009b, Aharonian et al. 2006).
The green curve in Fig. \ref{pulsarinterp} represents a typical {\it Fermi} pulsar spectrum (a power-law slope of $\Gamma=-1.8$ and an exponential cutoff at 2 GeV), and is here supposed to be steady along the orbit, hence reproduced in both INFC and SUPC panels of Fig. \ref{pulsarinterp}.

There is nothing special in the shape of the assumed pulsar spectrum when compared with all pulsars found by Fermi. However, the flux level of the green curve is here constrained by observations: On one hand, it cannot be much larger, since otherwise (being unaffected by the orbital motion) the INFC \ls\ spectrum would have been found at a larger flux level. On the other hand, one could have assumed a lower-flux pulsar contribution, but in that case the pulsar magnetosphere would become more and more irrelevant the less contributing it is at all phase ranges along the orbit, and a yet to be developed theoretical model would need to account for the whole range of observations, from GeV to TeV, with no magnetosphere contribution. Hence, the most interesting testing scenario  is whether it is possible (and useful to understand the observed phenomenology) to have comparably contributing pulsar and wind components when the GeV emission is maximum (equivalently, a pulsar domination when the GeV emission is minimum).

The light shaded area in Fig. \ref{pulsarinterp}, between 10--100 GeV, emphasizes the region where no data is available at the moment.
The other shadowed regions in Fig. \ref{pulsarinterp} represent the minimum and maximum of the wind component allowed by the GeV data, given the assumption of the pulsar contribution, at INFC and SUPC. At INFC (left panel, red shadow) the assumed pulsar spectrum is consistent with the data itself, hence the minimum allowed value of the wind component is compatible with zero. 
In both plots, the dashed lines represent the wind contribution spectra assumed to derive the red and blue solid lines. These solid lines are the sum up of the pulsar (green line) and the wind (dashed lines) components. 

One can see from Figure \ref{pulsarinterp} that the appearance of spectral cutoff {\it and} the GeV variability can be accommodated in the framework of this idea.
In particular, if these two components are emitting GeV photons such that, e.g., at the SUPC both contribute similarly to the GeV flux (see Figure \ref{pulsarinterp} right panel), then:

\begin{itemize}
\item The anti-correlation of GeV-TeV fluxes is naturally maintained, since it is a generic feature embedded in inverse Compton models describing the TeV fluxes and the pulsar only sums up to it an orbitally-steady contribution. 
Essentially: from inverse Compton models,  the GeV emission is enhanced (reduced) when the highly relativistic electrons seen by the observer encounter the seed photons head-on (rear-on),
e.g, see Boettcher \& Dermer (2005), Bednarek (2007). 

\item {\it Fermi} sees pulsar-like spectra which varies with the orbital phase as the resulting effect of a changing dominance of the two contributions along the orbit. For instance, 
 at ~10 GeV in INFC the
wind-related contribution, responsible of the TeV flux, may naturally take over and be larger than that of the
pulsar, leading to the disappearance of the cutoff.
The left panel of Fig. \ref{pulsarinterp} shows that if one sums up a
pulsed and a wind
contribution, the
cutoff may naturally disappear (or get very large) in INFC. In the case of SUPC (blue lines), instead, one is summing up a pulsed contribution with
something that also must increase (because of the predicted GeV-TeV anti-correlation
found in all inverse Compton models, and the constrain of the last {\it Fermi} upper limit) towards low energies and thus the cutoff is
maintained. A possible increase of the energy (or total disappearance) of the GeV measured exponential cutoff (i.e., why there is no cutoff in the INFC of \ls; and why we see such a high cutoff in the integrated-along-the-orbit spectrum of \lsi) can thus be accommodated within this idea. {\it Fermi} maybe summing up components that in its range of energies are changing dominance from the inner pulsar to the outer wind contributions.
 
\item It would  
be even harder to detect the pulsation in GeV data (since the pulsed emission is only a fraction of the $\gamma$-ray flux observed). However, this concept also implies that a pulsation search in {\it Fermi} data will be more successful in those phase ranges corresponding to the {\it Fermi} minimum (INFC), where the pulsed fraction is necessarily larger than at SUPC, due to the hardness of the spectrum measured at the neighboring energy band.

\item Pulsar models based on wind processes (intra or inter-wind) must underproduce the detected  {\it Fermi} emission, since they lack the contribution made directly by the pulsar magnetosphere.

\end{itemize}

We caveat though that we presented here only a   conceptual exploration of this possibility, hence the relative level between the contributions can be altered while maintaining the same overall idea.  
We also emphasize that the shape of the curves plotted in Fig. \ref{pulsarinterp} for the pulsar and wind contribution is very general. The wind contribution responsible for generating the TeV radiation should be
below the upper limit put by {\it Fermi} at 10 GeV; and, particularly at SUPC, it should contribute more at lower than at higher energies (if based on inverse Compton, e.g., Bottcher \& Dermer 2005, Dubus 2006, Sierpowska-Bartosik \& Torres 2008).

This 2-components idea can be tested by future observations. In particular, further observations by {\it Fermi} can confirm or rule out this concept by investigating the yellow energy range (as reported in Fig. \ref{pulsarinterp}) increasing the observing time.
Another possibility to test the 2 component model is with a very deep {\em INTEGRAL} exposure of LS 5039. The current   {\em INTEGRAL}  fluxes, derived from 3\,Ms of {\em IBIS/ISGRI} data  are  $(3.54 \pm 2.30)\times10^{-11}$~erg~cm$^{-2}$~s$^{-1}$ for INFC, and a flux upper limit for the SUPC phase interval of $1.45\times10^{-11}$~erg~cm$^{-2}$~s$^{-1}$ (90\% conf. level) in the 25--200\,keV energy band (Hoffmann et al. 2009). The pulsar spectrum is not expected to be variable with the orbit, and the slope in the hard X-ray should be the extrapolation of that in the MeV band (before the cut-off, of course). We verified that the extrapolation of the assumed pulsar contribution (green line in Fig. \ref{pulsarinterp}) to low energies was in agreement with these results, and especially  with the {\em INTEGRAL} upper limit at SUPC.  
More  {\em IBIS/ISGRI} data needs to be collected to possibly constrain the spectral slope of the putative pulsar in LS 5039.

In several recent conferences where results of the VERITAS array have been presented (e.g. Aliu et al. 2010, this volume), it was noted that \lsi\ could now be in a low TeV-state at the usual phases where it was detected at such energies; whereas, it seems to be shining normally (or even at a higher level) in GeV. \lsi\ is a more complex source than \ls, and Be-wind variations can introduce variability in the orbital profiles. The latter has been already found in X-rays (Torres et al. 2010).  With data now at hand, and lacking orbitally-resolved GeV and TeV follow up, we cannot say more as to the impact of the 2-components idea in the case of \lsi.

Finally, we remark that we are not ruling out that a comprehensive model of the wind shock or wind zone contribution  that is producing the TeV radiation may --on its own-- be enough to describe the GeV phenomenology in detail. We can only safely say that the current models did not appropriately predict it, despite the level of detail they already show. On the other hand, it can only be pertinent to consider that if these sources are pulsar-composed (see, e. g., Dubus 2006a), their magnetospheric contribution could be comparable to that found in all other pulsars detected by the {\it Fermi} LAT. If so, this idea presents a natural setup to understand the observed variability in the GeV spectra of these sources. 

\section{Epilogue}

At the moment of writing, new VHE facilities are in the process of design. The CTA/AGIS class would put together an array of dozens of IACTs in order to enhance the sensitivity (aimed to be improved  by one order of magnitude at 1 TeV when compared with H.E.S.S./VERITAS) and energy acceptance (from a few tens of GeV to 100 TeV). One of the aspects that such facilities could study is 
the formation of relativistic outflows from highly magnetized, rotating objects.
For instance, data on LS 5039, analyzed in the context of PWZ models already rule out that mono-energetic electrons are responsible for most of the emission. What else can future data tell us? How would that impact on current models for dissipation in pulsar winds (see, e.g., Jaroschek et al. 2008 and references therein)? It is not implausible that close systems may trigger different phenomenology within the PWZ, ultimately affecting particle acceleration there. Models of particle energization and dissipation in pulsar winds are currently made for isolated objects, and we lack knowledge on whether the inclusion of such objects in close binaries will affect the wind behavior or even the  magnetosphere in appreciable ways. Could we gain knowledge on this using future short-timescales  $\gamma$-ray  observations?
Also  among the possibilities for future instruments, it is also worth noticing that of the determination of  the duty cycles of high-energy phenomena by using
continued observations of key objects (such as Cyg X-1) with current instrument's sensitivity using sub-arrays of future ACTs.
It is interesting to note that neutrino detection or non-detection with ICECUBE will also shed light on the nature of the $\gamma$-ray emission / limit the neutrino-to-photon ratio irrespective of the system composition (e.g.,  see Aharonian et al. 2006b, and Torres \& Halzen 2007 for applications to LS 5039 and \lsi, respectively). For instance, Neronov \& Ribordy (2009) presented a hadronic model for $\gamma$-ray binaries in which the multi-TeV neutrino flux from the source can be much higher and/or harder than the detected TeV flux, and where most neutrinos are produced in pp interactions close to the bright massive star, in a region optically thick for the TeV photons. The secondary pairs of these processes would participate in the emission at lower energies. 
The nature of LS 5039 and \lsi\ $\gamma$-ray binary systems is still unknown. Whereas current data are consistent, and in some cases prefer, an interpretation based on non-accreting pulsars a final proof in either way (pulsations, accretion lines) is pending.\\

{\it This work has been supported by grants AYA2009-07391 and SGR2009-811. 
The author warmly acknowledges A. Caliandro, R. Dubois, G. Dubus, D. Hadasch, N. Rea, and A. Sierpowska-Bartosik for discussions.  }


\begin{thebibliography}{99}

\bibitem{} Abdo A. et al. 2009, ApJ Letters 701, 123 
\bibitem{} Abdo A. et al. 2009b, ApJ Letters 706, 56
\bibitem{} Abdo A. et al. 2009c, Science 326, 1512
\bibitem{} Abdo A. et al. 2010, ApJS 187, 460
\bibitem{} Aharonian F., et al. 2005a, A\&A 442, 1 
\bibitem{} Aharonian F., et al. 2005b, Science 309, 746 
\bibitem{} Aharonian F., et al. 2006, A\&A 460, 743 
\bibitem{} Aharonian F., et al. 2007, A\&A, 469, L1
\bibitem{} Aharonian F., Anchordoqui L. A., Khangulyan D., \& Montaruli T. 2006b, Journal of Physics: Conference Series 39, 408
\bibitem{} Acciari V. A. et al. 2008, ApJ 679, 1427 
\bibitem{} Acciari V. A. et al. 2009, ApJ 700, 1034 
\bibitem{} Albert J. et al. 2006, Science 312, 1771 
\bibitem{} Albert J. et al. 2007, ApJ Letters 665, 51 
\bibitem{} Albert J. et al. 2008, ApJ 684, 1351 
\bibitem{} Albert J. et al. 2009, ApJ 693, 303 
\bibitem{} Aleksi\'c J. et al. 2010, ApJ in press; arXiv: 1005.0740
\bibitem{} Aragona C. et al.  2009, ApJ 698, 514
\bibitem{} Arons J., \& Tavani M. 1993, ApJ, 403, 249
\bibitem{} Arons J., \& Tavani M. 1994, ApJS, 90, 797
\bibitem{} Atwood W. B., Ziegler M, Johnson R. P. and Baughman B. M. 2006, ApJ Letters 652, 49

\bibitem{} Ball L., \& Kirk J. G. 2000, Astroparticle Physics, 12, 335
\bibitem{} Bednarek W. 1997, A\&A 322, 523
\bibitem{} Bednarek W. 2006, MNRAS 368, 579 
\bibitem{} Bednarek W. 2007, A\&A 464, 259 
\bibitem{} Belczynski K. \& Ziolkowski J. 2009, arXiv:0907.4990
\bibitem{} Bignami G. F., Maraschi L., \& Treves A. 1977, A\&A 55, L155 
\bibitem{} Bogovalov S. V. \& Aharonian F. 2000, MNRAS 313, 504
\bibitem{} Bogovalov S. V. et al. 2008, MNRAS 387, 63
\bibitem{} Bosch-Ramon V., Paredes, J. M., Rib\'o, M., et al. 2005, ApJ, 628, 388
\bibitem{} Boettcher M. \& Dermer C. D. 2005, ApJ 634, 81




\bibitem{} Casares J., et al. 2005a, MNRAS, 364, 899 
\bibitem{} Casares J., et al. 2005b, MNRAS, 360, 1105 
\bibitem{} Cassinelli, J. P. 1979, ARA\&A 17, 275
\bibitem{} Chernyakova, M. et al. 2006, MNRAS, 372, 1585



\bibitem{}  Dhawan V., Mioduszewski, A., Rupen, M. 2006, VI Microquasar Workshop, PoS 52.1
\bibitem{}  Dubus G. 2006a, A\&A, 456, 801
\bibitem{} Dubus G. 2006b, A\&A 456, 80
\bibitem{} Dubus G., Cerutti B., \& Henri G. 2008, A\&A 477, 691





\bibitem{} Gregory P. C. 2002, ApJ 575, 427 
\bibitem{} Gregory P. C. et al. 1979, AJ 84, 1030 
\bibitem{} Gregory P. C., \& Neish C. 2002, ApJ, 580, 1133 
\bibitem{} Grundstrom E. D. et al. 2007, ApJ 656, 437



\bibitem{} Harrison F.~A., et al. 2000, ApJ 528, 454
\bibitem{} Hickox R. C., Narayan R., Kallman T. R, 2004, ApJ 614, 881
\bibitem{} Hinton J. A.,  et al. 2009, ApJ Letters 690, 101
\bibitem{} 	Hoffmann, A. D., Klochkov, D., Santangelo, A., Horns, D., Segreto, A., Staubert, R., \& P\"uhlhofer, G., 2009, A\&A 494, L37

\bibitem{} Israel G.L. \& Stella L. 1996, ApJ 468, 369


\bibitem{} Jaroschek C. H., et al. 2008, Advances in Space Research 41, 481
\bibitem{} Johnston S., et al. 1992, ApJ, 387, L37
\bibitem{} Johnston S. Manchester, R.N., McConnell D., Campbell-Wilson D., 1999, MNRAS 302, 277
\bibitem{} Johnston S., Ball L., Wang N., Manchester R. N., 2005 MNRAS, 358, 1069



\bibitem{} Kirk J. G., Ball, L., \& Skjaeraasen O. 1999, Astroparticle Physics, 10, 31



\bibitem{} Leahy D. A., Harrison F. A., \& Yoshida A. 1997, ApJ, 475, 823


\bibitem{} McSwain M. V., et al. 2004, ApJ, 600, 927 

\bibitem{} Maraschi L. \& Treves A. 1981, MNRAS, 194, 1
\bibitem{} Martocchia A., Motch C., Negueruela I. 2005, A\&A 430, 245
\bibitem{}  Massi M.,  Rib\'o M., Paredes J. M., Peracaula, M \& Estalella, M. 2001, A\&A 376, 217
\bibitem{}  Massi M.,  et al. 2004, A\&A 414, L1
\bibitem{} Mart\'i J. et al. 1998, A\&A 239, 951
\bibitem{} Mart\'i J. et al. 2004, A\&A 418, 271
\bibitem{} Melatos A. 1998 Memorie della Societa Astronomia Italiana 69, 1009



\bibitem{} Neronov A., \& Ribordy M. 2009, Phys. Rev. D79, 043013

\bibitem{} Paredes J. M., Mart\'i J., Rib\'o M., \& Massi, M. 2000, Science, 288, 2340 
\bibitem{} Paredes J. M., Rib\'o M., Ros E., Mart\'i J., \& Massi M. 2002, A\&A, 393, L99
\bibitem{} Paredes J. M., et al. 2007, ApJ 664, L39
\bibitem{} Protheroe R. J. \& Stanev T. 1987, ApJ 322, 838


\bibitem{} Rea N., Torres D. F., van der Klis M., Jonker P., Mendez M., \& Sierpowska-Bartosik A. 2010, MNRAS 405, 2206
\bibitem{} Reig P., et al. 2000, MNRAS 317, 205
\bibitem{} Rib\'o M., Reig, P., Mart\'i, J., \& Paredes, J. M. 1999, A\&A, 347, 518 
\bibitem{} Rib\'o M., Paredes, J. M., Romero, G. E., et al. 2002, A\&A, 384, 954
\bibitem{} Rib\'o M., Paredes, J. M., Mold\'on J., \& Mart\'i, J. 2008, A\&A Letters, arXiv: 0801.2940
\bibitem{} Romero G. E., Okazaki A. T., Orellana M., Owocki S. P., 2007,
A\&A 474, 15



\bibitem{} Sidoli L. et al. 2006, A\&A 459, 901 
\bibitem{} Sierpowska A. \& Bednarek W. 2005, MNRAS 356, 711
\bibitem{} Sierpowska-Bartosik A. \& Torres D. F. 2007, ApJ Letters 671, 145
\bibitem{} Sierpowska-Bartosik A. \& Torres D. F. 2008a,  ApJ Letters 674, 89
\bibitem{} Sierpowska-Bartosik A. \& Torres D. F. 2008b,  Astroparticle Physics 30, 239
  \bibitem{} Sierpowska-Bartosik A. \& Torres D. F. 2009,  ApJ 693, 1462





\bibitem{} Tavani M., \& Arons, J. 1997, ApJ, 477, 439
\bibitem{} Takahashi T., et al. 2009, ApJ, 697, 535
\bibitem{} Tauris T.~M., van den Heuvel E.~P.~J., 2006, in Lewin W. H. G., van der Klis, M., eds., Compact stellar X-ray sources, Cambridge University Press, Cambridge, p.\ 623 
\bibitem{} Torres D. F. \&  Halzen F. 2007, Astroparticle Physics 27, 500
  \bibitem{} Torres D. F., et al., 2010, ApJ Letters 719, 104

\bibitem{} Vaughan B. A., et al. 1994, ApJ 435, 362

\bibitem{} Waters L. B. F. M. 1986, A\&A 162, 121






\bibitem{} Zdziarski A. A., Neronov A., \& Chernyakova M. 2008, arXiv:0802.1174
\bibitem{} Zhang S., Torres D. F., Li J., Chen Y. Rea N, \& Wang J., MNRAS in press, arXiv 1006.1427




\end{thebibliography}
\end{document}